\documentclass[aps,prl,reprint,superscriptaddress]{revtex4-2}
\pdfoutput=1% Only for arXiv with pdf figures!
\usepackage{graphicx}% Include figure files
\usepackage{dcolumn}% Align table columns on decimal point
\usepackage[version=4]{mhchem}
\usepackage{gensymb}
\usepackage{xcolor}
\usepackage{hyperref}
\hypersetup{colorlinks,breaklinks,pdfusetitle=true}
\begin{document}
%\preprint{}

\title{Improved Measurements of Muonic Helium Ground-State Hyperfine Structure \\at a Near-Zero Magnetic Field}


\author{P.~Strasser}
  \email[]{patrick.strasser@kek.jp}
  %\thanks{}
  \affiliation{
    Muon Science Laboratory, 
    Institute of Materials Structure Science (IMSS), 
    High Energy Accelerator Research Organization (KEK), 
    1-1 Oho, Tsukuba, Ibaraki 305-0801, Japan}
  \affiliation{
    Muon Science Section, 
    Materials and Life Science Division, 
    J-PARC Center, 
    2-4 Shirakata, Tokai-mura, Naka-gun, Ibaraki 319-1195, Japan}
  \affiliation{
    Materials Structure Science Program, 
    Graduate Institute for Advanced Studies, 
    SOKENDAI, 
    1-1 Oho, Tsukuba, Ibaraki 305-0801, Japan}
%----------
\author{S.~Fukumura}
  \email[]{fukumura@phi.phys.nagoya-u.ac.jp}
  \affiliation{
    Department of Physics, 
    Nagoya University, 
    Furo-cho, Chikusa-ku, 
    Nagoya 464-8601, Japan}
%----------
\author{R.~Iwai}
  \affiliation{
    Muon Science Laboratory, 
    Institute of Materials Structure Science (IMSS), 
    High Energy Accelerator Research Organization (KEK), 
    1-1 Oho, Tsukuba, Ibaraki 305-0801, Japan}
%----------
\author{S.~Kanda}
  \affiliation{
    Muon Science Laboratory, 
    Institute of Materials Structure Science (IMSS), 
    High Energy Accelerator Research Organization (KEK), 
    1-1 Oho, Tsukuba, Ibaraki 305-0801, Japan}
  \affiliation{
    Muon Science Section, 
    Materials and Life Science Division, 
    J-PARC Center, 
    2-4 Shirakata, Tokai-mura, Naka-gun, Ibaraki 319-1195, Japan}
  \affiliation{
    Materials Structure Science Program, 
    Graduate Institute for Advanced Studies, 
    SOKENDAI, 
    1-1 Oho, Tsukuba, Ibaraki 305-0801, Japan}
%----------
\author{S.~Kawamura}
  \affiliation{
    Department of Physics, 
    Nagoya University, 
    Furo-cho, Chikusa-ku, 
    Nagoya 464-8601, Japan}
%----------
\author{M.~Kitaguchi}
  \affiliation{
    Department of Physics, 
    Nagoya University, 
    Furo-cho, Chikusa-ku, 
    Nagoya 464-8601, Japan}
  \affiliation{
    Kobayashi-Maskawa Institute, 
    Nagoya University, 
    Furo-cho, Chikusa-ku, 
    Nagoya 464-8602, Japan}
%----------
\author{S.~Nishimura}
  \affiliation{
    Muon Science Laboratory, 
    Institute of Materials Structure Science (IMSS), 
    High Energy Accelerator Research Organization (KEK), 
    1-1 Oho, Tsukuba, Ibaraki 305-0801, Japan}
  \affiliation{
    Muon Science Section, 
    Materials and Life Science Division, 
    J-PARC Center, 
    2-4 Shirakata, Tokai-mura, Naka-gun, Ibaraki 319-1195, Japan}
%----------
\author{S.~Seo}
  \affiliation{
    Graduate School of Arts and Sciences, 
    The University of Tokyo, 
    3-8-1 Komaba, Meguro, Tokyo 153-8902, Japan}
%----------
\author{H.~M.~Shimizu}
  \affiliation{
    Department of Physics, 
    Nagoya University, 
    Furo-cho, Chikusa-ku, 
    Nagoya 464-8601, Japan}
%----------
\author{K.~Shimomura}
  \affiliation{
    Muon Science Laboratory, 
    Institute of Materials Structure Science (IMSS), 
    High Energy Accelerator Research Organization (KEK), 
    1-1 Oho, Tsukuba, Ibaraki 305-0801, Japan}
  \affiliation{
    Muon Science Section, 
    Materials and Life Science Division, 
    J-PARC Center, 
    2-4 Shirakata, Tokai-mura, Naka-gun, Ibaraki 319-1195, Japan}
  \affiliation{
    Materials Structure Science Program, 
    Graduate Institute for Advanced Studies, 
    SOKENDAI, 
    1-1 Oho, Tsukuba, Ibaraki 305-0801, Japan}
%----------
\author{H.~Tada}
  \affiliation{
    Department of Physics, 
    Nagoya University, 
    Furo-cho, Chikusa-ku, 
    Nagoya 464-8601, Japan}
%----------
\author{H.~A.~Torii}
  \affiliation{
    School of Science, 
    The University of Tokyo, 
    7-3-1 Hongo, Bunkyo-ku, Tokyo 113-0033, Japan}
%----------

\collaboration{MuSEUM Collaboration}
\noaffiliation

%\date{\today}
\date{December 15, 2023}

\begin{abstract}
Muonic helium atom hyperfine structure (HFS) measurements are a sensitive tool to test the three-body atomic system and bound-state quantum electrodynamics theory, and determine fundamental constants of the negative muon magnetic moment and mass. The world's most intense pulsed negative muon beam at the Muon Science Facility of the Japan Proton Accelerator Research Complex allows improvement of previous measurements and testing further $CPT$ invariance by comparing the magnetic moments and masses of positive and negative muons (second-generation leptons). 
We report new ground-state HFS measurements of muonic helium-4 atoms at a near-zero magnetic field, performed for the first time using a small admixture of CH$_{4}$ as an electron donor to form neutral muonic helium atoms efficiently. 
Our analysis gives $\Delta\nu$ = 4464.980(20)~MHz (4.5~ppm), which is more precise than both previous measurements at weak and high fields. 
The muonium ground-state HFS was also measured under the same conditions to investigate the isotopic effect on the frequency shift due to the gas density dependence in He with CH$_{4}$ admixture and compared with previous studies. Muonium and muonic helium can be regarded as light and heavy hydrogen isotopes with an isotopic mass ratio of 36. No isotopic effect was observed within the current experimental precision. 
\end{abstract}

%\keywords{}

\maketitle

%%%\section{}
%%%\subsection{}
%%%\subsubsection{}

%==============================================================================
%Introduction \textcolor{blue}{}

Muonic helium is composed of a helium atom with one of its two electrons replaced by a negative muon ($\mu^{-}$). 
It is formed when negative muons are stopped in helium gas with a small admixture of foreign gas following a complicated process. First, the muon is captured by a helium atom in a high muonic orbit and quickly ejects both electrons via Auger transitions \cite{Kim1971}. At high pressure (a few atmospheres), the muon then quickly cascades down through radiative transitions to the muonic 1$s$ ground state ($\sim$400 times smaller than the electronic 1$s$ state in H) of the muonic helium ion [$\mu^{-}$$^{4}$He$^{++}$]$^{+}$ 
%within 1 ns \cite{Kim1971, Evseev1975}. 
on a timescale $<$~1~ns \cite{Kim1971, Evseev1975}. 
To form muonic helium atoms, i.e., the neutral $\mu^{-}$$e^{-}$$^{4}$He$^{++}$ system, a collision with a foreign gas atom acting as an electron donor (typically Xe \cite{Souder1980}, here CH$_{4}$, see below) is necessary. 
The muon is so closely bound to the helium nucleus that it nearly completely screens one proton charge producing a ``pseudonucleus'' with a positive effective charge and a magnetic moment nearly equal to that of a negative muon ($\mu_{\mu^{-}}$). 
Thus, muonic helium can be regarded as a heavy hydrogen isotope, similar to muonium, another hydrogenlike atom made of a bound state of a positive muon and an electron ($\mu^{+}e^{-}$), and forms with it the longest isotopic chain (mass ratio of 36). 
Muonic helium and muonium have been used to study extreme isotopic effects in chemical reaction rates and test fundamental theories of chemical kinetics \cite{Fleming2011}. 
Recently, following the spectroscopy measurements of the $2s-2p$ transition in muonic hydrogen (known as the proton radius puzzle) \cite{Antognoni2013} and muonic deuterium \cite{Pohl2016}, the Lamb shift was also measured in muonic helium-3 and helium-4 ions ($\mu^{3,4}$He$^{+}$) to determine the charge radius \cite{Krauth2021, Schuhmann2023arXiv}.

The ground-state hyperfine structure (HFS) in a muonic helium atom results from the interaction of the remaining electron and the negative muon magnetic moment $\mu_{\mu^{-}}$ and is almost equal to that of muonium but inverted. 
High-precision measurements of the muonium ground-state HFS are regarded as one of the most sensitive tools for testing quantum electrodynamics (QED) theory \cite{Eides2019} and determining fundamental constants of the positive muon magnetic moment $\mu_{\mu^{+}}$ and mass $m_{\mu^{+}}$ \cite{Liu1999}. 
New precision measurements are being carried out at the Japan Proton Accelerator Research Complex (J-PARC) by the \mbox{MuSEUM} Collaboration \cite{Kanda2021, Nishimura2021, Tanaka2021}. 
In muonic helium, the HFS interval is also sensitive to variations of basic physical constants and can be used to determine the fine structure constant $\alpha$ (quadratic dependence) and the negative muon and helium masses \cite{Frolov2012}. 
It can improve the negative muon mass \cite{Beltrami1986} (see conclusion), but difficultly challenge $\alpha$ \cite{Morel2020} or helium mass \cite{Sasidharan2023}. 
The same microwave magnetic resonance technique as with muonium can be used to measure the muonic helium HFS interval $\Delta\nu$ and the negative muon magnetic moment $\mu_{\mu^{-}}$ and mass $m_{\mu^{-}}$. 
The principle of the experiment is to produce polarized muonic helium atoms, irradiate them with microwaves suitable to flip the muon spin, and exploit the parity-violating muon decay (the electron from the weak decay $\mu^{-} \rightarrow e^{-} + \overline{\nu_{e}} + \nu_{\mu}$ is emitted preferentially in the direction antiparallel to the $\mu^{-}$ spin) to sample the muon spin-flip probability and measure the resonance curve.

Previous measurements performed in the 1980s were statistically limited \cite{Orth1980, Gardner1982}. This can be significantly improved using the world's most intense pulsed negative muon beam at the Muon Science Facility (MUSE) of J-PARC \cite{Strasser2018}. 
Another key factor is efficiently producing neutral muonic helium atoms, the prerequisite to measuring HFS. 
After muon capture, the $\mu^{4}$He$^{+}$ ion cannot capture an electron from neighboring He atoms because its electron binding energy is similar to H (13.6~eV). 
CH$_{4}$ was preferred to Xe used previously for the charge neutralization because Xe's large atomic number ($Z=54$) prevents efficient $\mu^{-}$ capture by He ($Z=2$). Since the relative capture probability is proportional to the nuclear charge ratio (Fermi-Teller $Z$~law \cite{knight1976}), this effect can be reduced by a factor of 5 using CH$_{4}$ (total charge $Z=10$). 
CH$_{4}$ has a similar ionization energy of 12.5~eV and gives a residual $\mu^{-}$ polarization of $\sim$5\% \cite{Arseneau2016} like Xe \cite{Souder1980}. 
The initial muon polarization ($\sim$100\%) is strongly reduced during the muon cascade process in He due to Auger transition and collisional Stark mixing \cite{Evseev1975}, making it more challenging to measure muonic helium HFS compared to muonium (50\% polarization).

We report improved ground-state HFS measurements of muonic helium-4 atoms at a near-zero magnetic field using CH$_{4}$ as an electron donor. 
Muonium HFS was also measured under the same conditions to investigate the isotopic effect on the frequency shift due to the gas density dependence in He with CH$_{4}$ admixture and compared with previous studies.

%==============================================================================
%Experiment

\begin{figure}[t]
    \includegraphics[width=1.0\linewidth,clip]{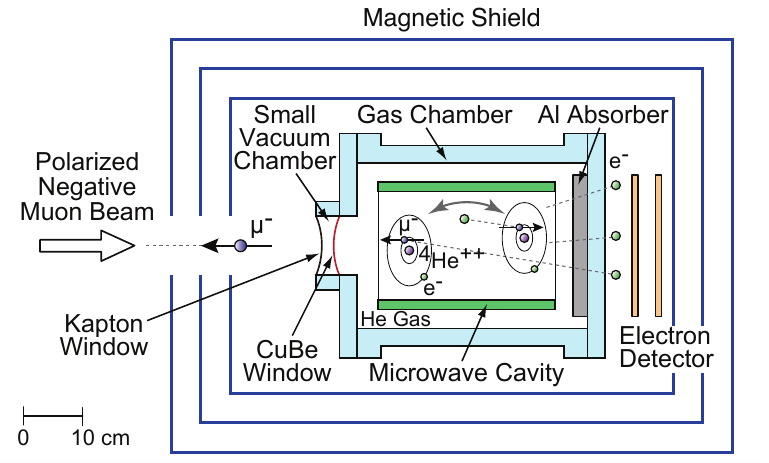}
    \caption{\label{Fig1}
    Schematic view of the experimental setup to measure ground-state muonic helium atom HFS at zero magnetic field. 
    }
\end{figure}

The experiment was performed at J-PARC MUSE \mbox{D-line} \cite{Higemoto2017} using the apparatus developed by the \mbox{MuSEUM} Collaboration to determine muonium HFS at zero field \cite{Kanda2021}. 
The experimental setup enclosed in a magnetic shield box made of three layers of permalloy is shown in Fig.~\ref{Fig1}. 
Pulsed polarized decay $\mu^{-}$ (backward decay, polarization $>$~90\%, double-pulsed structure 100~ns wide separated by 600~ns and repetitive at 25~Hz) were stopped into a microwave cavity placed inside an aluminum gas chamber containing pressurized helium gas with 2\% CH$_{4}$ admixture as an electron donor. 
Within a few nanoseconds, $\mu^{4}$He$^{+}$ ions were neutralized, a time short enough compared to the Rabi oscillations induced by the applied microwave.

%Fig1--------------------------------------------------------------------------
%------------------------------------------------------------------------------

The entrance beam window of the gas chamber was made of copper beryllium (CuBe) foil 10 cm in diameter. 
A small vacuum chamber with a 75~$\mu$m thick Kapton window was mounted on the entrance window to avoid deforming the CuBe foil while evacuating the gas chamber. 
The fiber hodoscope was removed (Fig.~1 in \cite{Kanda2021}). 
Three measurements were performed at an absolute He gas pressure of 3, 4, and 10.4~atm with a 50, 100, and 125~$\mu$m thick CuBe window, respectively. The muon beam momentum was tuned in each case to maximize the number of stopped $\mu^{-}$ in the microwave cavity with an optimum at 25, 27, and 30~MeV/$c$, respectively. The momentum spread of the beam was $\Delta p/p = 10\%$ (FWHM). 
The muon stopping rate and distribution in the cavity were estimated using a Monte Carlo simulation. 
Each measurement was performed under a user program in separate beam cycles with different primary proton beam power [typical beam intensity (0.7--1.2) 10$^{6}$~$\mu^{-}$/s].

The gas pressure was measured by a pressure transducer (Fluke RPM4 reference pressure monitor) with an accuracy of 0.02\%. 
The He/CH$_{4}$ gas mixture was prepared by filling the first 2\% of the nominal pressure with CH$_{4}$ gas followed by high-purity He gas with an accuracy of 0.2\%. 
The relative He/CH$_{4}$ ratio between measurements and the presence of other contaminants was confirmed by quadrupole-mass spectrometry sampling the gas through a capillary tube before and after the measurement.

The muon spin was flipped by applying a microwave magnetic field in the cavity. A larger cylindrical cavity (181~mm inner diameter, 304~mm long) developed to enable muonium HFS measurement at lower gas pressure without severe statistics loss was used \cite{Ueno2019}. The cavity resonates in TM220 mode with a tunable frequency range of 4462--4466~MHz and a quality factor of 11~400--11~700. 
The remaining microwave system was identical to that used in \cite{Kanda2021}. 
At the resonance, the microwave field induced the $\mu^{-}$ spin flip, changing the angular distribution of the electron ($e^{-}$) from the $\mu^{-}$ decay, which was detected with a segmented scintillation detector placed downstream. 
Measurements were performed by scanning the microwave frequency and measuring the electron asymmetry ($N_{\text{ON}}/N_{\text{OFF}}-1$) to determine the resonance frequency $\Delta\nu$. 
Since the cavity and the downstream beam stopper attached to the aluminum absorber (1 mm thick, not shown in Fig.~\ref{Fig1}) were all made of copper, $\mu$He signals were well separated from muonic copper ($\mu$Cu) background events due to different muon lifetimes. Mostly, $\mu$He signals remained by selecting delayed events while drastically reducing $\mu$Cu events.

%==============================================================================
%HFS Resonance Results

\begin{figure}[t]
    \includegraphics[width=1.0\linewidth,clip]{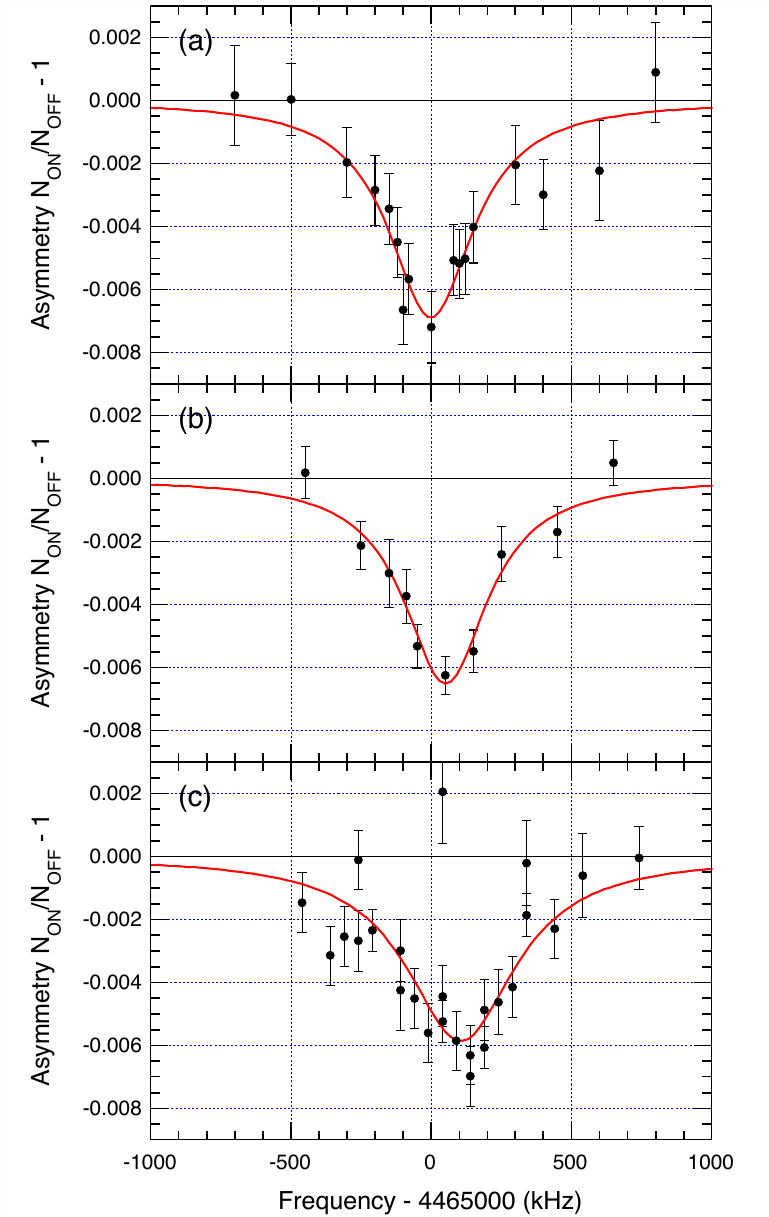}
    \caption{\label{Fig2}
    Muonic helium HFS resonance curve measured at zero field with He~+~CH$_{4}$(2\%) at (a) 3.0, (b) 4.0, and (c) 10.4~atm, respectively. Delayed events from 1.6~$\mu$s after the second muon pulse were selected. The solid lines represent the fitting results. 
    }
\end{figure}

Muonic helium HFS resonance curves measured chronologically at a He gas pressure of 4.0, 10.4, and 3.0~atm are shown in Fig.~\ref{Fig2} using delayed events from 1.6~$\mu$s after the second muon pulse. The data for these curves were obtained in 105, 63, and 76 h, respectively, including changing frequencies. 
The data analysis was performed by determining the hit cluster and taking coincidence between the two detector layers as described in \cite{Kanda2021}. Data with fluctuating microwave power feedback readings were ignored in the final analysis. 
The resonance curve centers were determined by fitting a theoretical resonance line shape using the ``{\it old muonium}'' method \cite{Liu1999} (same at zero field) from 1.6 to 60~$\mu$s. 
The reduced chi-squared values ($\chi^2$/ndf, where ndf is the number of degrees of freedom) in Fig.~\ref{Fig2} are 7.6/14, 5.5/7, and 47.0/22, respectively. 
The poor $\chi^2$ of Fig.~\ref{Fig2}(c) results from data taken near the resonance center with no Rabi-oscillation signals observed despite normal microwave power feedback readings.

%Fig2--------------------------------------------------------------------------
%------------------------------------------------------------------------------

During the last two measurements at 10.4 and 3.0~atm, a blind analysis method 
%\cite{Nishimura2023} 
was introduced that adds a \mbox{secret} offset value to the applied microwave frequency (randomly selected $\pm$8~kHz, fixed for all measurements), resulting in a measured resonance curve with that offset value. When the analysis was completed, the blind was opened, revealing the actual resonance frequency.

\begin{figure}[t]
    \includegraphics[width=1.0\linewidth,clip]{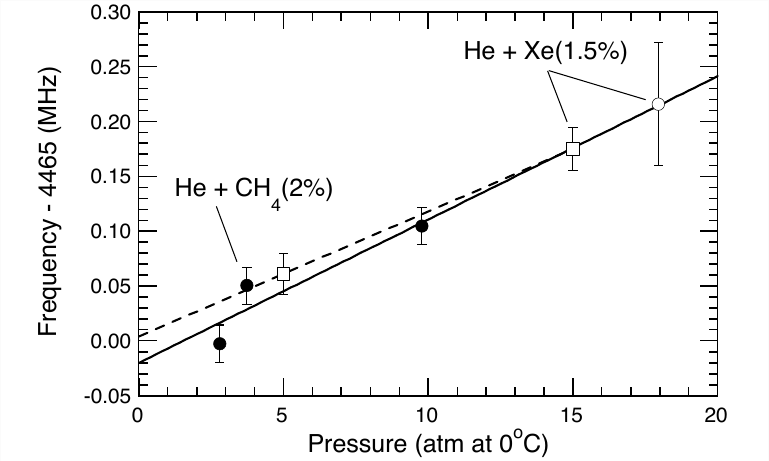}
    \caption{\label{Fig3}
    Muonic helium atom hyperfine frequency $\Delta\nu$ as a function of the He~+~CH$_{4}$(2\%) gas pressure (closed circle). The solid line shows the linear fit to the data to determine $\Delta\nu$(0). Previous results from \cite{Orth1980} (open circle) and \cite{Gardner1982} (open square) with the linear extrapolation (dashed line) measured with He~+~Xe(1.5\%) are also shown for comparison. 
    }
\end{figure}

The measured values for $\Delta\nu$ are shown in Fig.~\ref{Fig3} as a function of the gas density in atmosphere at 0$\celsius$ and corrected for nonideal gas behavior. Previous results from \cite{Orth1980, Gardner1982} measured with He~+~Xe(1.5\%) are also shown for comparison. 
The HFS frequency at zero pressure $\Delta\nu(0)$ of a free $\mu$He atom was obtained by fitting the data. 
It is known that hydrogenlike systems like muonium \cite{Casperson1975} and alkali atoms \cite{Dorenburg1979} show both a linear and quadratic pressure dependence on the gas buffer in which they are embedded. 
However, only the linear pressure shift coefficient resulting from competing short- and long-range interactions between the $\mu$He atom and the buffer gas at a given pressure \cite{Rao1970} can be obtained here. 
By fitting our measured values with $\Delta\nu(p) = \Delta\nu(0)~+~Ap$, we obtained $\Delta\nu$(0) = 4464.980(20)~MHz (4.5~ppm) and $A$ = 13.0~(3.2)~kHz/atm (0$\celsius$). 
The uncertainty indicated is mainly statistical. Systematic uncertainties 
for $\Delta\nu$(0) estimated to $<$~800~Hz are discussed later. 
Our measurement is consistent within  $1\sigma$ with previously obtained values of 4464.95(6)~MHz (13~ppm) at weak field \cite{Orth1980} and 4465.004(29)~MHz (6.5~ppm) at high field \cite{Gardner1982}. 
Although muonic helium is similar to muonium, the theoretical approach (reviewed in \cite{Krutov2008}) to determine $\Delta\nu$ has been limited due to the three-body interaction. Higher-order QED effects estimated to be around 130~ppm are still not yet fully considered \cite{Krutov2008, Pachucki2001, Aznabayev2018}.

%Fig3--------------------------------------------------------------------------
%------------------------------------------------------------------------------

%==============================================================================
%Pressure Shift

\begin{figure}[t]
    \includegraphics[width=1.0\linewidth,clip]{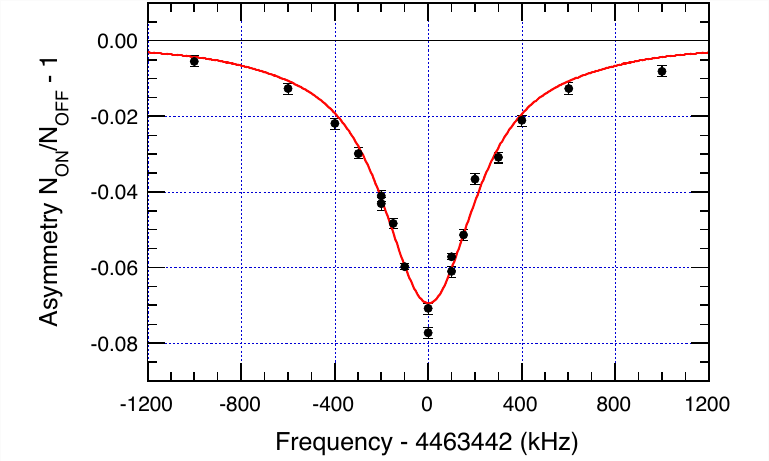}
    \caption{\label{Fig4}
    Muonium HFS resonance curve measured under the same experimental conditions of He~+~CH$_{4}$(2\%) at 10.4~atm. The same time selection as in Fig.~\ref{Fig2} was applied. 
    }
\end{figure}

The first observation of the muonic helium atom HFS resonance \cite{Orth1980} used the pressure shift correction from muonium, measured under the same conditions, to determine $\Delta\nu$(0) from only one measurement at 19.4~atm. This was justified since no isotopic effects were observed for muonium and H, D, and T in noble gases \cite{Morgan1973, Casperson1975}. 
The isotopic effect on the pressure shift in He with CH$_{4}$ admixture was investigated by measuring muonium HFS under the same conditions using decay $\mu^{+}$ (just reversing the polarity of the \mbox{D-line} magnets). Figure~\ref{Fig4} shows the muonium HFS resonance curve measured at 10.4~atm. The positron asymmetry is nearly ten times larger for muonium, which is consistent considering 50\% polarization for muonium as opposed to $\sim$5\% for muonic helium. 
Combining this muonium measurement $\Delta\nu_{}$(10.4~atm) = 4463.4382(23)~MHz with an earlier one at 4~atm (from a \mbox{MuSEUM} beamtime) and with $\Delta\nu_{}$(0) from \cite{Liu1999}, we obtain for muonium in He~+~CH$_{4}$(2\%) a linear pressure shift coefficient of 13.8(2)~kHz/atm (0$\celsius$). 
This value is consistent within the uncertainty to the value reported in \cite{Orth1980} measured with He~+~Xe(1.5\%).

Table \ref{Tab1} shows a comparison of the linear pressure shift coefficients for hydrogenlike atoms in He. 
The preliminary value for muonium in pure He \cite{Seo2019} was obtained indirectly from a study using Kr/He mixture to reduce the pressure shift effect in muonium HFS measurements. %(to be published). 
The hydrogen pressure shift data are shown for pure He \cite{Pipkin1962} and He+Xe(1.5\%) where the fractional pressure shift is calculated using the measured values from \cite{Pipkin1962, Ensberg1968} (1.5\% Xe reduces the linear pressure shift coefficient in He by nearly 8\%). 
Unfortunately, no pressure shift data were ever reported for light hydrogenlike atoms in CH$_{4}$; only for $^{133}$Cs atoms \cite{Beer1976} where the linear pressure shift is negative as in Xe, while it is positive in He. Thus, He with a small admixture of CH$_{4}$ is expected to behave similar to Xe and slightly reduce the total pressure shift. 
No isotopic effect can be seen within the current experimental precision comparing muonic helium with muonium. Also, admixtures of 2\% CH$_{4}$ and 1.5\% Xe seem to have similar effects. 
More precise measurements with $\mu$He atoms would be needed to confirm the tendency that heavier atoms have slightly larger pressure shifts, as was suggested for tritium and hydrogen in Ne and Ar (no data for T in He) \cite{Pipkin1962}.

%Fig4--------------------------------------------------------------------------
%------------------------------------------------------------------------------

%Tab1--------------------------------------------------------------------------
\begin{table}[b]
\caption{\label{Tab1}
Comparison of the linear pressure shift coefficients for hydrogenlike atoms in He (unit: kHz/atm at 0$\celsius$).}
\begin{ruledtabular}
\begin{tabular}{clll}
\textrm{Atom}&
\multicolumn{1}{c}{\textrm{He+CH$_4$(2\%)}}&
\multicolumn{1}{c}{\textrm{He+Xe(1.5\%)}}&
\multicolumn{1}{c}{\textrm{Pure He}}\\
&\multicolumn{1}{c}{\it this work}&&\\
\colrule
Muonium & 13.8 $\pm$ 0.2 & 14.7 $\pm$ 0.9 \cite{Orth1980}               & 17.0 $\pm$ 1.6 \cite{Seo2019}\\
H       &                              & 15.0 $\pm$ 0.3 \cite{Pipkin1962,Ensberg1968} & 16.3 $\pm$ 0.3 \cite{Pipkin1962}\\
$\mu$He  & 13.0 $\pm$ 3.2 & 11.4 $\pm$ 2.7 \cite{Gardner1982} &   \\
\end{tabular}
\end{ruledtabular}
\end{table}
%------------------------------------------------------------------------------

%==============================================================================
%Systematics

The systematic uncertainties of the current experiment are shown in Table \ref{Tab2} for $\Delta\nu$(0). Other uncertainties are common with \cite{Kanda2021}. 
The detector pileup is negligible due to decay $\mu^{-}/\mu^{+}$ intensities being 10 times smaller compared to surface ($\mu^{+}$) muons used in \cite{Kanda2021}. 
The error on the pressure depends essentially on the temperature uncertainty when converting to 0$\celsius$ and can be reduced to about $\sim$5~Hz with better temperature control by keeping fluctuation below 0.1$\celsius$. 
Following the approach described in \cite{GardnerPhD1983}, the upper-limit effect of the quadratic terms $B$ on $\Delta\nu$(0) [i.e., $\Delta\nu(p)=\Delta\nu(0)+Ap+Bp^2$], which results from the three-body interaction of a muonic helium atom with two gas buffer atoms \cite{Ray1972}, was estimated using the most precise measurement of $B$ for muonium in Kr \cite{Casperson1975}. This is justified by the fact that $B$ becomes smaller as the atomic number of the noble gas decreases and appears isotope independent \cite{Dorenburg1979}. 
With $B$ as a fixed value (nonzero), we obtain an upper limit of $\delta\Delta\nu$(0)~=~780~Hz. Additional high-pressure measurements would allow the determination of $B$. 
The effect on the CH$_{4}$ concentration is more difficult to ascertain because of the unknown value of its pressure shift. 
As an upper value, assuming a shift for CH$_{4}$ similar to the largest known value for H in Xe \cite{Ensberg1968}, the present concentration accuracy of 0.2\% corresponds to an uncertainty of $\sim$3~Hz/atm. This can be reduced by using the same mixture from a gas container for all measurements.

%Tab2--------------------------------------------------------------------------
\begin{table}[]
\caption{\label{Tab2}
Systematic uncertainties in the experiment.}
\begin{ruledtabular}
\begin{tabular}{lc}
\multicolumn{1}{l}{\textrm{Source}}&
\multicolumn{1}{c}{\textrm{Contribution}}\\
\colrule
Pressure gauge precision (Hz) & 5\\
Gas temperature fluctuation (Hz) & 45\\
CH$_{4}$ concentration (Hz/atm) & 3\\
Quadratic pressure shift (Hz) & $<$ 780\\
Detector pileup (Hz) & Negligible\\
\end{tabular}
\end{ruledtabular}
\end{table}
%------------------------------------------------------------------------------

%==============================================================================
%Summary and Prospects

After nearly 40 years, new precise measurements of the muonic helium HFS were performed using the high-intensity pulsed negative muon beam at J-PARC MUSE. 
The result obtained at zero field with an uncertainty of 4.5~ppm has 3 times better precision than the previous direct measurement \cite{Orth1980} without relying on muonium data to determine the HFS frequency at zero pressure. It is also more precise than the indirect measurement at high field \cite{Gardner1982} and the first performed with CH$_{4}$ admixture to form neutral muonic helium atoms. 
Muonic helium HFS is the only available experimental data for three-body muonic atoms, but the latest theoretical value \cite{Aznabayev2018} is still 30 times less precise. 
However, recent groundbreaking theoretical calculations developed for HFS in $^3$He \cite{Patkos2023} could be applied to muonic helium HFS to improve the current theory to the same level as the present experimental accuracy, giving the first opportunity to test QED effects in three-body muonic atoms.

Muonium HFS measured under the same conditions does not reveal any isotopic effect on the linear pressure shift with muonic helium atoms within the current experimental precision. 
Muonium pressure dependence will not be used to determine $\Delta\nu(0)$, but further measurements with muonic helium at higher pressure are planned to measure the quadratic pressure coefficient.

High-field measurements are in preparation at the \mbox{H-line} \cite{Higemoto2017} after muonium HFS measurements, using 10 times more muon beam intensity than at the \mbox{D-line}, 
1~MW primary proton beam power, increased detection efficiency of decay electrons being more focused on the upstream and downstream detectors by the high-magnetic field, 
and utilizing the Rabi-oscillation spectroscopy method \cite{Nishimura2021}, 
aiming at a precision for $\Delta\nu$ below 100 ppb after 100 days. 
This will also permit one to determine $\mu_{\mu^{-}}$ and $m_{\mu^{-}}$ 
%This will also permit the determination of the negative muon magnetic moment and mass 
below 1~ppm to test $CPT$ invariance by comparison with positive muons. 
The ratio $\mu_{\mu^{-}}$/$\mu_{p}$ was previously determined to 47~ppm \cite{Gardner1982}. 
Several new measurements of the positive muon mass are now in progress \cite{Kanda2021, Uetake2019, Crivelli2018}. 
Presently, $\mu_{\mu^{+}}$ and $\mu_{\mu^{-}}$ provide a $CPT$ test at a level of 3~ppm \cite{Fei1994}, derived from the negative muon mass precision \cite{Beltrami1986}. 
A more precise $\mu_{\mu^{-}}$/$\mu_{p}$ is also needed to determine the negative muon magnetic moment anomaly $a_{\mu^{-}}$ in the muon $g-2$ experiment at Brookhaven National Laboratory. 
The more accurate $\mu_{\mu^{+}}$/$\mu_{p}$ \cite{Liu1999} is currently used for both $a_{\mu^{+}}$ and $a_{\mu^{-}}$ to test the standard model's predictions and $CPT$ theorem \cite{Bennett2004, Bennett2006}. 

Moreover, a new experimental approach to recover the polarization lost during the muon cascade is being investigated by repolarizing $\mu$He atoms using a spin-exchange optical pumping technique \cite{Barton1993}, which could further improve the measurement precision by nearly 1 order of magnitude, reaching ultimately $\mathcal{O}$(10 ppb) for $\Delta\nu$ and $\mathcal{O}$(100 ppb) for $\mu_{\mu^{-}}$.

\begin{acknowledgments}
The muon experiment at the Materials and Life Science Experimental Facility (MLF) of J-PARC was performed under a user program (Proposals No.~2020B0333, No.~2021B0169, No.~2022A0159). The authors would like to acknowledge the help and expertise of the staff of J-PARC MLF MUSE and thank K.~Shimizu, T.~Tanaka, H.~Yamauchi, H.~Yasuda, and F.~Yoshizu for their support in preparing and participating in the experiment. This work was supported by the JSPS KAKENHI Grant No. 21H04481. 
\end{acknowledgments}

\bibliography{muHeHFS2023}
\end{document}